\documentstyle[prb,aps,preprint]{revtex}

\begin{document}

\draft

\title{Crossover effects in the Wolf--Villain model of epitaxial growth
in 1+1 and 2+1 dimensions}
\author{Pavel \v{S}milauer\cite{byline1}}
\address{Interdisciplinary Research Centre for Semiconductor Materials,
Imperial College, London SW7~2BZ, United Kingdom}

\author{Miroslav Kotrla}
\address{Institute of Physics, Czech Academy of Sciences, Na Slovance 2,
180~40 Praha 8, Czech Republic}

\date{\today}

\maketitle

\begin{abstract}

A simple model of epitaxial growth proposed by Wolf and Villain is
investigated using extensive computer simulations. We find an
unexpectedly complex crossover behavior of the original model in both 1+1
and 2+1 dimensions. A crossover from the effective growth exponent
$\beta_{\rm eff}\!\approx\!0.37$ to $\beta_{\rm eff}\!\approx\!0.33$ is
observed in 1+1 dimensions, whereas additional crossovers, which we
believe are to the scaling behavior of an Edwards--Wilkinson type, are
observed in both 1+1 and 2+1 dimensions. Anomalous scaling due to
power--law growth of the average step height is found in 1+1 D, and also
at short time and length scales in 2+1~D. The roughness exponents
$\zeta_{\rm eff}^{\rm c}$ obtained from the height--height correlation
functions in 1+1~D ($\approx\!3/4$) and 2+1~D ($\approx\!2/3$) cannot be
simultaneously explained by any of the continuum equations proposed so
far to describe epitaxial growth.

\end{abstract}


Simple (``toy'') models of surface growth have become very popular
recently, mainly as examples of externally driven non-equilibrium systems
that exhibit simple (yet nontrivial) scaling behavior.\cite{review} The
focus is on the surface width (roughness) $w$ defined as
$w\!=\!\langle\sqrt{\overline{h^2}-\overline{h}^{2}}\rangle$ (the
overbars denote spatial averages, $\langle\rangle$ a statistical
average), {\it i.e.\/}, the variance of the surface height profile
$h({\bf r},t)$. The surface width scales with the time $t$ and the
linear dimension of the substrate $L$ as $w(t,L)\propto L^{\zeta}
f(t/L^{z})$, where the scaling function $f(x)$ has the properties
$f(x)\!=\!$ const, $x\!\gg\!1$ and $f(x)\!\propto\!x^{\beta},
x\!\ll\!1$, $\beta\!=\!\zeta/z$. Thus, $w$ grows according to a power
law, $w \propto t^{\beta}$, until a steady state characterized by a
constant value of the width is reached after a time $t_{sat}$
proportional to $L^z$. The value of the saturated width
$w_{sat}$ varies with the system size according to $w_{sat} \propto
L^{\zeta}$. The exponents $\zeta$ and $\beta$ (or $\zeta$ and $z$)
characterize the scaling behavior of the roughness for a particular
model and determine its universality class in analogy with theory of
critical phenomena.

Alternatively, one can study the surface roughness using the
height--height correlation function $G({\bf r},t)\!=\!\langle [h({\bf
x}\!+\!{\bf r},t)\!-\!h({\bf x},t)]^2\rangle$ which obeys the scaling
relation\cite{review} $G(r,t)\!\propto\!r^{2\zeta^{\rm c}}
g(r/t^{1/z^{\rm c}})$, where the scaling function $g(x)$ is constant for
$ x\!\ll\!1$ and $g(x)\!\propto\!x^{-2 \zeta^{\rm c}}$ for $x\!\gg\!1$
(equivalently, the structure factor $S$ can be used, see e.g.
Ref.~\onlinecite{pr}). In most of growth models the exponents obtained
using the two different methods are equal.\cite{review}

Scaling behavior of the surface roughness can be investigated using
continuum, stochastic differential equations which in the case of
conserving models with surface diffusion have the form

\begin{equation}
\frac{\partial h({\bf r},t)}{\partial t}= -\nabla \cdot {\bf
j}({\bf r},t)+\eta ({\bf r},t)~~, \label{eq:1}
\end{equation}

\noindent where $\eta({\bf r},t)$ is a zero mean, random noise term in
the incoming flux, and the current ${\bf j}({\bf r},t)$ is a function of
the derivatives of $h({\bf r},t)$. In a number of recent theoretical
studies,\cite{vill,wv,st,ls,kls,sg,wvz,kps,ssw} models in which
surface diffusion is the dominant physical mechanism of the surface
smoothing were studied. The scaling relation $2\zeta\!=\!z-d^\prime$
(where $d^{\prime}$ is the substrate dimension) holds for these
models.\cite{wv} The most often studied cases were ${\bf
j}\propto -\nabla h$ [Edwards--Wilkinson (EW) model\cite{ew}], ${\bf
j}\propto\nabla\nabla^2 h$ (the {\it linear\/} diffusion
model\cite{vill,mull}), and ${\bf j}\propto \nabla(\nabla h)^2$ and
${\bf j}\propto (\nabla h)^3$ (the {\it nonlinear\/} diffusion
models\cite{wv,ls,ssw} which we will denote I and II, respectively). The
predicted values of exponents are\cite{vill,wv,ls}
$\beta^{EW}\!=\!(3-d)/4$, $\zeta^{EW}\!=\!(3-d)/2$,
$\beta^{lin}\!=\!(5-d)/8$, $\zeta^{lin}\!=\!(5-d)/2$,
$\beta^{nonlin-I}\!=\!(5-d)/(7+d)$,  $\zeta^{nonlin-I}\!=\!(5-d)/3$, and
$\beta^{nonlin-II}\!=\!(5-d)/2(3+d)$,  $\zeta^{nonlin-II}\!=\!(5-d)/4$
where $d\!=\!d^{\prime}+1$. It should be noted that these values are
usually based on renormalization group (RG) calculations within the
one--loop approximation (e.g., Refs.~\onlinecite{ls,tn}) or on
Flory--type approximation (Ref.~\onlinecite{vill}).

Alternative approach is to employ a powerful computer and study discrete
models with microscopic rules reflecting physically important surface
processes.\cite{conn} From a wide variety of discrete models we focus
our attention on the one proposed by Wolf and Villain\cite{wv} (WV) which
is based on the solid--on--solid model of crystal growth\cite{sos} and has
been designed with growth by molecular--beam epitaxy (MBE) in mind. The
growth rules of the model (see below) are supposed to mimic surface
diffusion, the principal mechanism of surface smoothing during MBE.
Original simulations of the model\cite{wv} in 1+1~D yielded exponents
$\beta_{\rm eff}\!=\!0.365\!\pm\! 0.015$ and  $\zeta_{\rm
eff}\!=\!1.4\!\pm\! 0.1$ (thus $z_{\rm eff}\!=\!3.8\!\pm\! 0.5$) in
agreement with the theoretical prediction of the {\it linear\/} model.
However, a subsequent numerical work\cite{kls} has shown that in 2+1~D
the values of the exponents are $\beta_{\rm eff}\!=\!0.206\!\pm\! 0.02$
and $\zeta_{\rm eff}\!=\!0.66\!\pm\! 0.03$ (thus $z_{\rm
eff}\!=\!3.2\!\pm\! 0.5$) which correspond to the prediction of the {\it
nonlinear\/} model I. The puzzling difference between the behavior of the
model in 1+1 and 2+1~D has been confirmed in another numerical
study.\cite{sg} The possible explanation could be a slow crossover from
exponent of the linear model to the exponent of the nonlinear model I
observable in 1+1 D only for large system sizes and long times, similar
to a crossover observed in a full--diffusion solid--on--solid
model.\cite{wvz} To complicate the situation even further, it has been
proposed recently by Krug, Plischke, and Siegert\cite{kps} that the WV
model should cross over to EW behavior for more than $\approx 10^{6}$
(1+1~D) or $\approx 2\times10^{4}$ (2+1~D) deposited layers. This result
was based on the study of the inclination--dependent diffusion current
which is supposed to generate the EW term ($\nabla^2h$) in continuum
differential equations.\cite{conn,cont} The EW term is more relevant (in
the RG sense) than all allowed nonlinear terms\cite{tn} and governs the
asymptotical behavior of the model. The long time needed to observe the
asymptotic regime may be explained as due to a very small coefficient in
front of the EW term. Finally, it has been found very recently that the
WV model in 1+1~D does not fulfill standard scaling and that different
values of  the exponents are obtained from behavior of the surface width
and the correlation function or the structure factor.\cite{ssw} The
exponents obtained in 1+1~D from the behavior of the structure factor
were $\zeta^{\rm c}_{\rm eff}\!=\!0.75\!\pm\! 0.05$ and $z^{\rm c}_{\rm
eff}\!=\!2.4\!\pm\! 0.1$ (thus $\beta^{\rm c}_{\rm eff}\!\approx\!0.31$)
very close to the nonlinear model II predictions.\cite{ssw} The authors
also suggested that this anomalous behavior is present in 2+1~D as well.

The need for better simulation data apparent from the above summary has
motivated the present work. We performed large--scale simulations of the
WV model in 1+1 and 2+1~D.  In 1+1~D, we used system sizes as large as
$L\!=\!40\,000$ sites and deposited up to $2^{27}\approx 1.3\times10^8$
layers whereas in 2+1~D we used lattice sizes of up to $1000\times 1000$
depositing up to $2^{17}\approx 1.3\times 10^5$ layers. Our results  for
exponents obtained from the surface width show (i) a crossover from
$\beta_{\rm eff}\approx 0.37$ ($\beta^{lin}$) to $\beta_{\rm eff}\approx
0.33$ ($\beta^{nonlin-I}$) in 1+1~D, (ii) crossovers from $\beta_{\rm
eff} \approx \beta^{nonlin-I}$  to the scaling behavior which we believe
is of the EW model in both 1+1 and 2+1~D. Exponents calculated from the
correlation function are in agreement with those of the nonlinear model
II in 1+1 D ($\zeta^{\rm c}_{\rm eff}\approx 0.75$) and the
nonlinear model I in 2+1 D ($\zeta^{\rm c}_{\rm eff}\approx 0.65$).

The microscopic rules of the basic model are the same as in Ref.
\onlinecite{wv}. In every time step, a particle is added at a randomly
chosen lattice site and then relaxes toward a nearest--neighbor site which
offers the highest {\it coordination\/} (the number of nearest neighbors)
where it sticks for the rest of the simulation. If the number of nearest
neighbors cannot be increased, particularly in the case of {\it tie\/}
(one or more neighboring sites have the same coordination as the original
site) the particle stays at the initial position. The generalization to
2+1~D is straightforward.

In 1+1~D, we carried out simulations for lattice sizes $L\!=\!$150, 300,
600, 800, 1000, 2000, and $40\,000$ depositing from $2^{25}$ to $2^{27}$
layers (except for $L\!=\!40\,000$ where only $2^{23}$ layers were
deposited), see Fig.~\ref{fig_1d}. (Note that the curves for lattice
sizes $L\!\ge\!800$ have been offset in Fig.~\ref{fig_1d} to avoid
confusion due to overlapping data points; statistical errors of the data
points are not larger than the symbols size.) The crossover of the
exponent $\beta_{\rm eff}$ from the value $0.372 \pm 0.007$ corresponding
to the linear model ($\beta^{lin}\!=\!\frac{3}{8}$) to $\approx 0.336 \pm
0.009$ in very good agreement with the nonlinear model I prediction
($\beta^{nonlin-I}\!=\!\frac{1}{3}$) can be best observed for the largest
simulated lattice, $L\!=\!40\,000$. It is apparent from Fig.~\ref{fig_1d}
that this crossover takes place after approximately $2^{15} \approx
3\times 10^4$ layers have been deposited. The values of $w_{sat}$ we have
obtained ($48\pm 1, 91\pm 10$, and $209\pm 20$ for $L\!=\!150, 300$, and
600, respectively), give $\zeta_{\rm eff}\!=\!1.05\!\pm\!0.1$, in very
good agreement with the nonlinear model I prediction, $\zeta_{\rm
eff}^{nonlin-I}\!=\!1$.

However, a more surprising observation can be made after close inspection
of the data in Fig.~\ref{fig_1d}. All the curves (with the possible
exception of the one for $L\!=\!40\,000$ where fewer layers were
deposited) show another decrease in the value of the exponent $\beta_{\rm
eff}$ after the deposition of $2^{21}\!\approx\!2\times 10^6$ layers. The
curves for $L\!\ge\!800$ seem to follow the slope corresponding to the
value $\beta\!=\!\frac{1}{4}$ of the EW model in 1+1~D (see
Fig.~\ref{fig_1d}). Because the time $t_{\rm sat}(L)$ needed for the
saturation depends on the system size as $L^z$, we obtain $t_{\rm
sat}(L\!=\!2000) > 2^{27}$ based on our results for $t_{\rm sat}$ for
small lattice sizes (even if we use $z^{nonlin-II}\!=\!5/2$). Moreover,
the position of the crossover seems not to depend on the lattice size.
Hence we can exclude that this is the effect of the saturation.

Similar behavior is revealed by an inspection of 2+1~D results in
Fig.~\ref{fig_2d}. The data given for the lattice sizes $L\!=\!500$ and
$L\!=\!1000$, correspond at first (between $\approx 2^3$ and $2^{10}$
layers deposited) to the value of $\beta_{\rm eff}\!=\!0.22 \pm 0.05$,
very close to (but definitely higher than) the nonlinear model I
prediction $\frac{1}{5}$. After $\approx 10^4$ layers are
deposited, a crossover takes place. Again it cannot be due to the
vicinity of the saturation regime because the estimate of $t_{sat}$ in
$2+1~D$ for $L\!=\!500$ is larger than $2^{22}$ layers based on our
previous results\cite{kls} and the position of the crossover does not
shift with the lattice size. A logarithmic increase of the roughness
($\beta^{EW}\!=\!0$) is expected for the EW model in 2+1~D.
Double--logarithmic and semilogarithmic plots of the last seven data
points (averaged over the both curves) are shown in the inset of
Fig.~\ref{fig_2d}. The statistics of the data are unsufficient to draw a
definitive conclusion, but it seems that the semilogarithmic plot
follows a straight line whereas the log--log one is curved. Thus, we
believe our data suggest that in both 1+1 and 2+1~D we observe crossovers
to the scaling behavior of the EW model predicted by Krug {\it et
al.\/}\cite{kps} and it should be noted that even the positions of these
crossovers agree well with this prediction.

Following the paper by Schroeder {\it et al.\/}\cite{ssw} we also
studied the behavior of $G(r,t)$. The time dependence of $G(r,t)$ in 1+1
and 2+1~D is shown in Fig.~\ref{fig_cf}. The exponent $\zeta^{\rm
c}_{\rm eff}\!=\!0.75\!\pm\!0.03$ we obtained in 1+1~D using the initial
slope of the correlation function for $t\!=\!2^{19}$ monolayers and
$L\!=\!40\,000$ (Fig.~\ref{fig_cf} (a)) differs substantially from the
value $\zeta_{\rm eff}\!\approx\!1$ obtained from the surface width
behavior at later times (see above) and corresponds to
$\zeta^{nonlin-II}\!=\!3/4$. However, the value $\zeta^{\rm c}_{\rm
eff}\!=\!0.65\!\pm\!0.03$ obtained in 2+1~D (Fig.~\ref{fig_cf} (b)) for
$L\!=\!500$ and $t\!=\!2^{17}$ monolayers is definitely different from
2+1~D value of $\zeta^{nonlin-II}\!=\!1/2$ and is much closer to
$\zeta^{nonlin-I}\!=\!2/3$ which was also obtained from the surface width
behavior.\cite{kls} The values of the exponents $\zeta^{\rm c}_{\rm eff}$
obtained from $G(r,t)$ are thus to the best of our knowledge inconsistent
with any simple model of epitaxial growth proposed so far.

In agreement with Ref.~\onlinecite{ssw}, we see the initial power--law
increase of the average step height $G(1,t)\!\propto\!t^\lambda$ with
$\lambda\!=\!0.38\!\pm\!0.01$ in 1+1~D, and
$\lambda\!=\!0.095\!\pm\!0.003$ in 2+1 D, see the insets in
Fig.~\ref{fig_cf}. In Ref.~\onlinecite{ssw}, it is shown how this can
explain effective exponents corresponding to the linear equation and the
discrepancy between $\zeta_{\rm eff}$ and $\zeta^{\rm c}_{\rm eff}$ in
1+1 D provided the nonlinear model II is used. Notice also that the
crossover from the exponent $\beta_{\rm eff}\!\approx\!0.375$ to the
value $\beta_{\rm eff}\!\approx\!0.33$ coincides with the beginning of
$G(1,t)$ saturation. In 2+1 D, if we use the values of the {\it nonlinear
model I\/} and  $\lambda\!\approx\!0.095$, we obtain $\zeta_{\rm
eff}\!\approx\!0.84$ and $\beta_{\rm eff}\!\approx\!0.23$. The value of
$\beta_{\rm eff}$ is in agreement with the results obtained here for
short times $t\!\leq\!2^{10}$ [see above; notice again that this is the
region where we observe the power--law growth of $G(1,t)$] and also the
value of $\zeta_{\rm eff}$ is not inconsistent with the initial slope of
the plot $w$ vs. $L$ in Fig. 2 of Ref.~\onlinecite{kls}. Hence, the
theory of Ref.~\onlinecite{ssw} agrees with our results provided we use
different ``underlying'' models in 1+1 and 2+1~D. The anomalous
behavior due to power--law growth of $G(1,t)$ is much weaker in 2+1 D
but can be observed at short time and length scales.

We also tried to find changes in the correlation function behavior that
should take place following the crossovers which we suppose are to the EW
behavior (see the thick lines in Fig.~\ref{fig_cf}). In 1+1 D
[Fig.~\ref{fig_cf} (a)], a decrease to the value of $\zeta^{\rm c}_{\rm
eff}\!\approx\!2/3$ is observed for $L\!=\!2000$ and $2^{27}$ monolayers
deposited, but this is still significantly higher than the EW model value
$\zeta^{EW}\!=\!1/2$. In 2+1 D [Fig.~\ref{fig_cf} (b)], the behavior of
the correlation function after the crossover does not change appreciably.
A probable explanation is that the range of length scales
(larger than the relevant crossover length) where $G(r,t)$ is influenced
by the EW behavior is too short to be resolved in our data. However, we
have found that the structure factor calculated for $L\!=\!2000$ and
$2^{27}$ monolayers does at large wavelengths follow the slope $-2$
expected for the EW behavior. Statistics of these data are not
sufficient to provide unambiguous proof and more work is needed.

In conclusion, we have studied kinetic roughening in 1+1 and 2+1
dimensions of a simple model of epitaxial growth proposed by Wolf and
Villain.\cite{wv} From the study of the surface width in 1+1~D, we have
found a crossover from the scaling behavior of a linear
differential equation proposed to describe MBE growth ($\beta_{\rm
eff}\!\approx\!0.37$) to that corresponding to a nonlinear model
I equation ($\beta_{\rm eff}\!\approx\!0.33$). According to
Ref.~\onlinecite{ssw} and our results, unusual behavior of the model in
1+1 D is due to a power--law growth of the average step height,
$G(1,t)\!\propto\!t^\lambda$ where $\lambda\!\approx\!0.38$ in 1+1 D and
$\approx\!0.095$ in 2+1 D. As a consequence, the exponent $\zeta^{\rm
c}_{\rm eff}$ obtained in 1+1~D from the height--height correlation
function differs from the value obtained from the scaling of the surface
width and does not change during the crossover. Its value in 1+1~D
($\zeta^{\rm c}_{\rm eff}\!\approx\!3/4$) corresponds to another
nonlinear model II proposed by Lai and Das Sarma.\cite{ls} However, the
value of the exponent $\zeta^{\rm c}_{\rm eff}\!\approx\!2/3$ in 2+1 D is
consistent with the nonlinear model I. One possibility of how to
explain the discrepancy between the behavior in 1+1 and 2+1 D in terms
of a continuum equation is to suppose that both relevant
nonlinearities [$\nabla^2(\nabla h)^2$ and $\nabla(\nabla h)^3$] are
present but with different coefficients (and thus different importance)
in 1+1 and 2+1~D.

In both 1+1 and 2+1~D, we see additional crossovers and our results seem
to indicate these are to the scaling behavior of the EW model. The
conclusion that the asymptotic behavior of the ``ideal'' MBE growth model
is of EW type is in agreement with the recent suggestion by Krug {\it et
al.\/}\cite{kps}. It would be interesting to see this crossover also in
other discrete models, in particular in the full--diffusion model of
Ref.~\onlinecite{wvz} which was very successful in describing the initial
stages of MBE growth. In some variants of the models with surface
diffusion, the final crossover to the EW behavior may be absent and
instead the instability may develop. We expect this kind of behavior in
models with barriers to interlayer transport.

We thank Professor D. Wolf and Professor D.D. Vvedensky for stimulating
discussions and communicating their results prior to publication, and to
Dr. M. Wilby for a critical reading of the first version of the
manuscript. The calculations reported here were performed on a Meiko
Infinity Series i860 Computing Surface at Imperial College and on Cray
Y/MP-EL at the Institute of Physics. Work performed at Imperial College
was supported by Imperial College and the Research Development
Corporation of Japan. The work was supported in part by grant No. 110110
of the Academy of Sciences of Czech Republic.

\begin{figure}
\caption{Surface width \it vs.\/ \rm time for the WV model in 1+1 D
($L\!=\!150 (\triangle), 300 (\times), 600$
(\protect\raisebox{1mm}{\protect\framebox[7pt]{~}}), $800 (\circ), 1000
(\bullet), 2000 (\diamond),$ and $40\,000$
(\protect\raisebox{0.5mm}{\protect\rule{3mm}{0.8mm}})). Notice that data
for larger lattice sizes ($L\!\ge\!800$) were offset to avoid overlapping
of data points.}
\label{fig_1d}
\end{figure}

\begin{figure}
\caption{Surface width \it vs.\/ \rm time for the WV model in 2+1~D
($L\!=\!500 (\triangle)$ and $1000 (\circ)$). Data for $L\!=\!1000$ were
offset to avoid overlapping of data points. The inset shows the last
seven data points for $L\!=\!1000$ in semilogarithmic ($\bullet$) and
double logarithmic ($\circ$) scales (see text).}
\label{fig_2d}
\end{figure}

\begin{figure}
\caption{(a) The height--height correlation function $G(r,t)$ in 1+1~D
for $L\!=\!40\,000$ and the times $t\!=\!2^{1} (\circ), 2^{4} (\bullet),
2^{7} (\diamond), 2^{10} (\times), 2^{13}$
(\protect\raisebox{1mm}{\protect\framebox[7pt]{~}}), $2^{16} (\ast),
2^{19} (\triangle), 2^{22} (+),$ and for $L\!=\!2000$ and $t\!=\!2^{27}$
(\protect\raisebox{0.5mm}{\protect\rule{3mm}{0.8mm}}); (b) The
height--height correlation function $G(r,t)$ in 2+1~D for $L\!=\!500$
and the times $t\!=\!2^2 (\ast), 2^{4}$
(\protect\raisebox{1mm}{\protect\framebox[7pt]{~}}), $2^{6} (\times),
2^{8} (\triangle), 2^{10} (\bullet), 2^{12} (\circ), 2^{14} (\diamond)$,
and $2^{17}$ (\protect\raisebox{0.5mm}{\protect\rule{3mm}{0.8mm}}). The
insets show the power--law increase of $G(1,t)$.}
\label{fig_cf}
\end{figure}


\begin{references}

\bibitem[*]{byline1} Also at: Department of Physics, The Blackett
Laboratory, Imperial College, London SW7~2BZ, United Kingdom; on leave
from Institute of Physics, Cukrovarnick\'{a} 10, 162~00~Praha~6, Czech
Republic

\bibitem{review} For reviews, see, e.g., F. Family, Physica A {\bf 168},
561 (1990); J. Krug and H. Spohn, in {\em Solids Far from Equilibrium:
Growth, Morphology and Defects}, edited by C. Godr\`{e}che, (Cambridge
University Press, Cambridge, 1990), p. 479; M. Kotrla, Czech. J. Phys. B
{\bf 42}, 449 (1992).

\bibitem{pr} M. Plischke, Z. R\'{a}cz, and D. Liu, Phys. Rev. B {\bf
35}, 3485 (1987).

\bibitem{vill} J. Villain, J. Phys. I {\bf 1}, 19 (1991).

\bibitem{wv} D.E. Wolf and J. Villain, Europhys. Lett. {\bf 13}, 389
(1990).

\bibitem{st} S. Das Sarma and P. Tamborenea, Phys. Rev. Lett. {\bf 66},
325 (1991).

\bibitem{ls} Z.-W. Lai, S. Das Sarma, Phys. Rev. Lett. {\bf 66}, 2348
(1991).

\bibitem{kls} M. Kotrla, A.C. Levi and P. \v{S}milauer, Europhys. Lett.
{\bf 20}, 25 (1992).

\bibitem{sg} S. Das Sarma and S.V. Ghaisas, Phys. Rev. Lett. {\bf 69},
3762 (1992). See also M. Plischke, J.D. Shore, M. Schroeder, M. Siegert,
and D.E. Wolf, {\it ibid.\/} {\bf 71}, 2509 (1993) and S. Das Sarma
and S.V. Ghaisas, {\it ibid.\/} {\bf 71}, 2510 (1993).

\bibitem{wvz} M.R. Wilby, D.D. Vvedensky, and A. Zangwill, Phys. Rev. B
{\bf 46}, 12896 (1992); (errata) {\it ibid.\/} {\bf 47}, 16068 (1993).

\bibitem{kps} J. Krug, M. Plischke, and M. Siegert, Phys. Rev. Lett.
{\bf 70}, 3271 (1993); (errata) {\it ibid.\/} {\bf 71}, 949 (1993).

\bibitem{ssw} M.~Schroeder, M.~Siegert, D.E.~Wolf, J.D.~Shore, and
M.~Plischke, Europhys. Lett. {\bf 24}, 563 (1993)..

\bibitem{ew} S.F. Edwards and D.R. Wilkinson, Proc. Roy. Soc. London
Ser. A {\bf 381}, 17 (1982).

\bibitem{mull} W.W. Mullins, J. Appl. Phys. {\bf 28}, 333 (1957).

\bibitem{tn} L.--H. Tang and T. Natterman, Phys. Rev. Lett. {\bf 66},
2899 (1991).

\bibitem{conn} In many cases, it is very difficult (or maybe impossible)
to establish an explicit connection between a specific microscopic model
and a continuum differential equation. In fact, we know about the only
systematic attempt to obtain a continuum equation from the rules of a
discrete model [D.D. Vvedensky, A. Zangwill, C. Luse, and M.R. Wilby,
Phys. Rev. E {\bf 48}, 852 (1993)] but even this method employs several
approximations. In this paper, we compare the results from simulations of
the WV model with results from various continuum models proposed to
describe MBE growth as it is usual in literature but the reader should
keep the above caveat in mind.

\bibitem{sos} G.H. Gilmer and P. Bennema, J. Appl. Phys. {\bf 43},
1347 (1972); J.D. Weeks and G.H. Gilmer, Adv. Chem. Phys. {\bf  40}, 157
(1979).

\bibitem{cont} The procedure described in the paper cited in
Ref.~\onlinecite{conn} does not provide the EW term for the WV model (just
the ordinary nonlinear equation proposed in Refs.~\onlinecite{wv,ls})
[D.D. Vvedensky (private communication)].

\end{references}
\end{document}